\magnification 1200
\centerline {\bf  Macroscopic Observables, Thermodynamic 
Completeness and the}
\vskip 0.2cm
\centerline {\bf Differentiability of Entropy}
\vskip 0.3cm
\centerline {{\bf  by Geoffrey L. Sewell}\footnote *{email: 
g.l.sewell@qmul.ac.uk}}
\vskip 0.3cm
\centerline {\bf School of Physics and Astronomy, Queen Mary 
University of London}
\vskip 0.2cm
\centerline {\bf Mile End Road, London E1 4NS, UK}
\vskip 0.5cm
\centerline {\bf Abstract}
\vskip 0.3cm
We provide a quantum statistical basis for (a) a characterisation of a 
complete set of of thermodynamic variables and (b) the differentiability 
of the entropy function of these variables. 
\vfill\eject
\centerline {\bf 1. Introduction}
\vskip 0.3cm
The fundamental formula of classical thermodynamics is
$$TdS=dE+pdV+{\sum}_{j=1}^{n}y_{j}dQ_{j},\eqno(1.1)$$ 
where $S$ is the entropy of a system, $T$ is its temperature, $p$ its 
pressure, $V$ its volume, $E$ its internal energy, $Q_{1},. \ .,Q_{n}$ a 
further set of conserved macroscopic variables and $y_{1},. \ .,y_{n}$ a 
set of real valued parameters representing the coupling of the system to 
external sources such that ${\sum}_{j=1}^{n}y_{j}dQ_{j}$ represents 
the work done by the system against these sources due to infinitesimal 
changes in the  $Q_{j}$\rq s. Thus, for fixed values of the $y_{j}$\rq s, 
the effective energy of the system  is 
$$E^{\rm (eff)}=E+{\sum}_{j=1}^{n}y_{j}Q_{j}.\eqno(1.2)$$
\vskip 0.2cm
The formula (1.1) implies that $S$ is a differentiable function of the 
variables $V, \ E$ and the $Q_{j}$\rq s, though it does not provide any  
general rule for identifying the latter variables of a system of given 
microscopic constitution. Thus we are confronted by the challenging 
problems of providing a quantum statistical basis for both a 
characterisation of the thermodynamic variables $(Q_{1},. \ .,Q_{n})$ 
and the differentiability of entropy. These are the problems that we shall 
address in this paper.
\vskip 0.2cm
Our treatment is formulated within the operator algebraic framework of 
quantum statistical mechanics wherein the model of a macroscopic 
system is represented as an infinitely extended one, as in the 
thermodynamic limit, and is designed to be applicable both to lattice 
systems and continuous ones The treatment is centred on (a) a set of 
macroscopic observables ${\hat q}=({\hat q}_{0}, \ {\hat q}_{1},. \ 
.,{\hat q}_{n})$ that are global densities of quantum versions of the 
above variables $(E, \ Q_{1}, \ .,Q_{n})$, and (b) the spatially ergodic 
states, i.e. the extremal elements of the convex set of translationally 
invariant states, which correspond to pure phases [1]. The equilibrium 
condition on these states is assumed to be given by the maximisation of 
the entropy density, subject to the constraint that fixes the expectation 
value of  ${\hat q}$ to a prescribed value $q$. The set ${\hat q}$ of 
macroscopic observables is then termed {\it thermodynamically 
complete} if, for each realisable value of $q$, there is precisely one 
spatially ergodic state that satisfies this condition. In other words, the 
values of the macro-observables ${\hat q}$ completely determine the 
pure phases of the model. Further, we find that it follows from the 
structure of the model that thermodynamic completeness implies that the 
resultant state of the model satisfies the Kubo-Martin-Schwinger (KMS) 
condition, whereby the dynamical automorphisms are the modular 
ones\footnote*{Recall that the general definition of the modulars is that 
they constitute a one-parameter group 
${\lbrace}{\sigma}_{t}{\vert}t{\in}{\bf R}{\rbrace}$ of automorphisms 
of a $W^{\star}$-algebra ${\cal M}$, induced by a faithful, normal state 
${\omega}$ according to the KMS condition that 
${\omega}([{\sigma}_{t}A]B)={\omega}(B{\sigma}_{t+i}A)$ for all 
$A,B$ in ${\cal M}$.}of the Tomita-Takesaki theory [2]. The 
differentiability of the entropy function then follows from the uniqueness 
of these automorphisms
\vskip 0.2cm
Our treatment proceeds as follows. In Section 2 we  provide a brief 
resume of classical thermodynamics in a form that connects simply with 
our subsequent formulation of the quantum model. In Section 3 we 
formulate that model in operator algebraic terms as an infinitely extended 
system. In Section 4 we formulate the equilibrium and thermodynamic 
completeness conditions discussed above In Section 5 we  show that, 
under the imposed conditions, the entropy function realised by the 
quantum model is indeed differentiable. We conclude in Section 6 with 
some brief observations about the basis of the treatment of quantum 
thermodynamics presented here.
\vskip 0.5cm
\centerline {\bf 2. The Classical Thermodynamic Model.}
\vskip 0.3cm
The classical phenomenological model is based on Eq. (1.1), which may 
be conveniently re-expressed in terms of the variables $Q=(Q_{0}:=E, \ 
Q_{1}, \ .,Q_{n})$ and ${\theta}=({\theta}_{0}, \ {\theta}_{1}, \ . 
,{\theta}_{n})$, where ${\theta}_{0}=T^{-1}$ and ${\theta}_{j}=T^{-
1}y_{j}$ for $j=1, \ .,n$. Eq.(1), as expressed in terms of these variables, 
then takes the form 
$$dS={\theta}.dQ+{\phi}dV,\eqno(2.1)$$
where
$${\phi}=T^{-1}p,\eqno(2.2)$$
which is termed the {\it reduced pressure}, and where the dot denotes the 
${\bf R}^{n+1}$ scalar product, i.e. 
${\theta}.dQ:={\sum}_{k=0}^{n}{\theta}_{k}dQ_{k}$. The 
components ${\theta}_{k}$ of ${\theta}$ are termed the {\it control 
variables}. 
\vskip 0.2cm
It is standard to classical thermodynamics [3] that 
\vskip 0.2cm\noindent
(a) the variables $V, \ S$ and $Q$ are extensive, while $T, \ {\phi}$ and 
${\theta} $ are intensive; and 
\vskip 0.2cm\noindent
(b) by the demand of thermodynamic stability, the entropy $S$ is a 
concave function of $Q$ and $V$. 
\vskip 0.2cm\noindent
It follows from (a) that $Q$ and $S$ take the forms $Vq$ and $Vs(q)$, 
where $q:=(q_{0}, \ q_{1}, \ .q_{n}), \ q_{j}=Q_{j}/V$ and $s$ is a 
function of $q$ only.. We term $q$ the {\it thermodynamic state} of the 
model. It follows from these specifications and Eq. (2.1) that 
$$d\bigl(s(q)V\bigr)={\theta}.dqV+{\phi}dV,$$
i.e. that 
$$(s(q)-{\theta}.q-{\phi})dV+(ds-{\theta}.dq)V=0,$$
which signifies that 
$$ds={\theta}.dq\eqno(2.3)$$
and
$${\phi}=s-{\theta}.q.\eqno(2.4)$$
We assume that the domains of definition of the variables $q$ and 
${\theta}$ are convex subsets of ${\bf R}^{(n+1)}$, which we denote by 
${\cal Q}$ and ${\Theta}$ and term the {\it thermodynamic state space} 
and the 
{\it control space}, respectively: quantum statistical specifications of 
these states will be provided in Section 4.  Since, by (b), $S$ is a concave 
function of its arguments, so too is its density, $s$. 
\vskip 0.2cm
Since Eq. (2.3) signifies that ${\theta}=s^{\prime}(q)$, the differential 
(i.e. the ${\bf R}^{(n+1)}$ gradient) of $s$, it follows that, for fixed 
${\theta},\  q$ is a stationary point of the function $s-{\theta}.(.)$ on 
${\cal Q}$. Moreover, as the concavity of $s$ ensures that the tangent to 
the graph of $s$ at $q$ lies above that graph, it follows that 
$$s(q^{\prime})-s(q){\leq}{\theta}.(q^{\prime}-q) \ {\forall} \ 
q^{\prime} {\in} \ {\cal Q},$$
i.e.
$$s(q)-{\theta}.q{\geq}s(q^{\prime})-{\theta}.q^{\prime} \ {\forall} \ 
q^{\prime}{\in}{\cal Q},\eqno(2.5)$$
which signifies that the stationary point $q$ of $s-{\theta}.(.)$ is a 
maximum. Hence, by Eq. (2.4) the reduced pressure ${\phi}$ is the 
Legendre transform of the entropy density $s$, i.e. 
$${\phi}({\theta})={\rm max}_{q{\in}{\cal Q}}\bigl(s(q)-
{\theta}.q\bigr),\eqno(2.6)$$
which implies, by the following argument, that ${\phi}$ is convex. For 
${\theta}, \ {\theta}^{\prime}{\in}{\Theta}, \ {\lambda}{\in}(0,1)$ and 
${\epsilon}>0$, it follows from Eq. (2.6) that, for some $q{\in}{\cal Q}$, 
$${\phi}\bigl({\lambda}{\theta}+(1-{\lambda}){\theta}^{\prime}\bigr)  \ 
<s(q)-\bigl({\lambda}{\theta}+(1-
{\lambda}){\theta}^{\prime}\bigr)+{\epsilon} \ {\equiv}$$  
$${\lambda}(s(q)-{\theta}.q)+(1-{\lambda})(s(q)-
{\theta}^{\prime}.q)+{\epsilon}{\leq}
{\lambda}{\phi}({\theta})+(1-
{\lambda}){\phi}({\theta}^{\prime})+{\epsilon}$$
and therefore, as ${\epsilon}$ is arbitrarily chosen,
$${\phi}\bigl({\lambda}{\theta}+(1-{\lambda}){\theta}^{\prime}\bigr)
{\leq}{\lambda}{\phi}({\theta})+
(1-{\lambda}){\phi}({\theta}^{\prime}),$$
which signifies that ${\phi}$ is convex.
\vskip 0.2cm
To summarise, the general structure of classical thermodynamics is 
governed by the forms of  the entropy function $s$ and the reduced 
pressure ${\phi}$, which is its Legendre transform. Further, $s$ is 
concave, ${\phi}$ is convex and the thermodynamic state $q$ is 
determined by the maximisation $s-{\theta}.(.)$. 
\vskip 0.5cm
\centerline {\bf 3. The Quantum Model} 
\vskip 0.3cm
In a standard way [1, 4-8]., we represent the quantum model of a 
macroscopic system, ${\Sigma}$, as an infinitely extended body: this 
corresponds to a treatment of the system in the thermodynamic limit. 
Specifically,we take the model to comprise a system of interacting 
particles that occupies an infinitely extended space, $X$, which may be 
either a Euclidean space, ${\bf R}^{d}$, or a lattice, ${\bf Z}^{d}$, with 
$d$ finite. Correspondingly, space translations are represented by $X$, 
considered as an additive group.
\vskip 0.3cm
{\bf The Algebra ${\cal A}$.}The model of ${\Sigma}$ is centred on a 
$C^{\star}$-algebra, ${\cal A}$, constructed in the following standard 
way. We denote by $L$ the set of bounded open subregions 
${\Lambda}$ of $X$. For each ${\Lambda}{\in}L$ we then construct, 
according to the general rules of quantum mechanics, a model 
${\Sigma}({\Lambda})$ of a finite system of particles of  the given 
species confined to ${\Lambda}$. Thus, the bounded observables of this 
system are represented by the self-adjoint elements of a type I primary 
$W^{\star}$-algebra ${\cal A}({\Lambda})$ in a separable Hilbert space  
${\cal H}({\Lambda})$, The constructions of the algebras ${\cal 
A}({\Lambda})$ is effected so as to meet the canonical requirements of 
isotony and local commutativity, i.e.
$${\cal A}({\Lambda}_{1}){\subset}{\cal A}({\Lambda}_{2}) \ 
{\rm and} \ {\cal H}({\Lambda}_{1}){\subset}{\cal H}({\Lambda}_{2}) 
\ {\rm if} \ {\Lambda}_{1}{\subset}{\Lambda}_{2}$$
and
$${\cal A}({\Lambda}_{1}){\subset}
{\cal A}({\Lambda}_{2})^{\prime} \ {\rm if} \ {\Lambda}_{1}{\cap}
{\Lambda}_{2}={\emptyset}$$
where the prime denotes commutant. Thus, 
${\cup}_{{\Lambda}{\in}L}{\cal A}({\Lambda})$ is a normed 
$^{\star}$-algebra, ${\cal A}_{L}$, of the local observables of the 
system ${\Sigma}$. We define ${\cal A}$ to be its norm completion.  
Thus 
${\cal A}_{L}$ and ${\cal A}$ are naturally identified as the algebras of 
local and quasi-local bounded observables, respectively, of  ${\Sigma}$. 
Its unbounded local observables are represented by the unbounded self-
adjoint operators affiliated to ${\cal A}_{L}$ [9]. In particular, the local 
Hamiltonian operator, $H({\Lambda})$, is the energy observable for the 
region ${\Lambda}$, and either belongs or is affiliated to  ${\cal 
A}({\Lambda})$.
\vskip 0.3cm
{\bf The Extensive Conserved Observables.} We assume that each 
${\Lambda}$ in $L$ harbours a set of intercommuting, linearly 
independent, possibly unbounded, extensive observables 
${\hat Q}({\Lambda})=\bigl(Q_{0}({\Lambda}),\ 
{\hat Q}_{1}({\Lambda}), \ .,{\hat Q}_{n}({\Lambda})\bigr),$ 
with ${\hat Q}_{0}({\Lambda})$ the internal energy $H({\Lambda}) $. 
These are designed to be the thermodynamic observables that correspond, 
for large ${\Lambda}$, to the classical variables $Q$ of Section 2. Their 
extensivity condition is that
$${\hat Q}({\Lambda}{\cup}{\Lambda}^{\prime})=
{\hat Q}({\Lambda})+ {\hat Q}({\Lambda}^{\prime}) \ {\rm if} \ 
{\Lambda}{\cap}{\Lambda}^{\prime}={\emptyset}.$$
\vskip 0.3cm
{\bf The Space Translational Automorphisms.} We represent space 
translations by a homomorphism, ${\sigma}$, of the group $X$ into the 
automorphisms of ${\cal A}$ satisfying the covariance conditions
$${\sigma}(x){\cal A}({\Lambda})={\cal A}({\Lambda}+x)
\eqno(3.1)$$
and
$${\sigma}(x){\hat Q}({\Lambda})={\hat Q}({\Lambda}+x).
\eqno(3.2)$$
\vskip0.2cm
We define the effective energy observable, 
$H_{\theta}^{\rm eff}({\Lambda})$, for the region ${\Lambda}$ to be 
the canonical counterpart to that given by Eq. (1.2) for the 
phenomenological model. Thus, by our definition of ${\theta}$ and its 
components ${\theta}_{j}$ in terms of $y_{1}, \ ,y_{n}$ and $T$,
$$H_{\theta}^{\rm eff}({\Lambda})=
{\theta}_{0}^{-1}{\theta}.{\hat Q}({\Lambda}).\eqno(3.3)$$
\vskip 0.3cm 
{\bf The State Space ${\cal S}$.} We take this to  comprise the positive 
normalised linear functionals on ${\cal A}$ whose restrictions to the 
local algebras ${\cal A}({\Lambda})$ are normal: this last condition is 
needed, in the case of continuous systems, to ensure that the number of 
particles in each bounded region ${\Lambda}$ is finite [10]. Thus, the 
restriction, ${\rho}_{\Lambda}$, of a state ${\rho}$ to the region 
${\Lambda}$ is represented by a density matrix, which we also denote by 
${\rho}_{\Lambda}$, i.e. 
$${\rho}(A)=Tr_{{\cal H}({\Lambda})}({\rho}_{\Lambda}A) \ {\forall} 
\  A{\in}{\cal A}({\Lambda}), \ {\Lambda}{\in}L.$$
We denote by ${\cal S}_{X}$ the set of translationally invariant states of 
${\Sigma}$, as defined by the condition that 
${\rho}={\rho}{\circ}{\sigma}(x)$ for all $x{\in}X$. Evidently ${\cal 
S}_{X}$ is convex. We denote the set of its extremal elements by ${\cal 
E}_{X}$. These are the spatially ergodic states of the system, 
characterised by the property that the space averages of the local 
observables are dispersionless in these states [1], i.e., defining the space 
average of a local observable $A$ to be
$${\tilde A}_{\Lambda}:={\vert}{\Lambda}{\vert}^{-1}
\int_{\Lambda}dx{\sigma}(x)A,$$ 
where ${\vert}{\Lambda}{\vert}$ is the volume of ${\Lambda}$; and 
for spatially ergodic ${\rho}$,
$${\rm lim}_{{\Lambda}^{\uparrow}X}
\bigl[{\rho}({\tilde A}_{\Lambda}^{2})-
{\rho}({\tilde A})^{2}\bigr]=0,$$
the limit being taken in the sense of Van Hove. 
\vskip 0.2cm
In particular, a {\it folium}, of states is defined [11] to be a norm closed 
convex subset, ${\cal F}$, of ${\cal S}$ that is stable under quasi-local 
modifications 
${\lbrace}{\rho}{\rightarrow}{\rho}(B^{\star}(.)B)/{\rho}(B^{\star}B
{\vert}B{\in}{\cal A}{\rbrace}$. The {\it normal folium} of a state 
${\rho}$ then corresponds to the set of  normal states on the GNS 
representation of ${\cal A}$ induced by ${\rho}$.
\vskip 0.3cm
{\bf Dynamics of the Model.}We assume that the dynamics of the finite 
system ${\Sigma}_{\Lambda}$ is given by the unitary transformations 
generated by the effective local Hamiltonian 
$H_{\theta}^{\rm eff}({\Lambda})$ on a time scale whose unit is 
${\theta}_{0}$. Thus, by Eq. (3.3),
$${\alpha}_{{\theta}{\Lambda}}(t)A=
{\rm exp}\bigl(i{\theta}.{\hat Q}({\Lambda})t\bigr)A
{\rm exp}\bigl(-i{\theta}.{\hat Q}({\Lambda})t\bigr) \ {\forall} \ 
t{\in}{\bf R}, \ A{\in}{\cal A}({\Lambda}).\eqno(3.4)$$ 
It follows from this formula and Eq. (3.1) that
$${\sigma}(x){\alpha}_{{\theta}{\Lambda}}(t)=
{\alpha}_{{\theta},{\Lambda}+x}(t){\sigma}(x).\eqno(3.5)$$
\vskip 0.2cm 
In order to formulate the dynamics of the infinitely extended system 
${\Sigma}$, we first recall that, in general, the model does not support an 
infinite volume {\it norm} limit of the automorphisms 
${\alpha}_{{\theta}{\Lambda}}$ (cf. [12, 13])\footnote*{To be more 
specific, such a limit is supported by spin systems with short range 
interactions [14-17 ], but not, in general, by continuous systems.}. To 
cope with this situation we formulate the dynamics of the model, in the 
Schroedinger representation [13], as a one-parameter group, 
${\lbrace}{\tau}(t){\vert}t{\in}{\bf R}{\rbrace}$, of normalised affine 
transformations of a folium ${\cal F}$ of its states that supports the 
infinite volume limit of the dual of  
${\alpha}_{{\theta}{\Lambda}}({\bf  R})$ given by the formula
$${\langle}{\tau}(t){\rho};A{\rangle}=
{\rm lim}_{{\Lambda}{\uparrow}X}{\langle}{\rho};
{\alpha}_{{\theta}{\Lambda}}(t)A{\rangle} \ {\forall} \ 
{\rho}{\in}{\cal F}, \ A {\in}{\cal A}_{L}, \  t{\in}{\bf R}.\eqno(3.6)$$
A consequence of this assumption is that [13]
$${\rm lim}_{{\Lambda}{\uparrow}X}
{\langle}{\rho};B_{1}[{\alpha}_{{\theta}{\Lambda}}(t)A]B_{2}
{\rangle}={\langle}{\hat {\rho}};{\pi}(B_{1})
[{\hat {\alpha}}_{\theta}(t){\pi}(A)]{\pi}(B_{2}){\rangle} \ 
{\forall} \ {\rho} {\in}{\cal F}, \ A, \ B_{1}, \ B_{2} \ {\in}
{\cal A}_{L}, \ t{\in}{\bf R},\eqno(3.7)$$
where ${\pi}$ is the GNS representation of the state ${\rho}, \ 
{\hat {\rho}}$ is its canonical extension to the $W^{\star}$-algebra 
${\cal A}_{\cal F}:={\pi}({\cal A})^{{\prime}{\prime}}$ and 
${\lbrace}{\hat {\alpha}}_{\theta}(t){\vert}t{\in}{\bf R}{\rbrace}$ is the 
one parameter group of automorphisms of that algebra defined by the 
formula
$${\hat {\alpha}}_{\theta}(t){\pi}(A)=
s-{\rm lim}_{{\Lambda}{\uparrow}X}
{\pi}\bigl({\hat {\alpha}}_{{\theta}{\Lambda}}(t)A\bigr) \ {\forall} \ 
A{\in}{\cal A}_{L}, \ t{\in}{\bf R}.\eqno(3.8)$$
where the limit is taken over an increasing sequence of cubes with 
parallel axes. Thus, the automorphisms ${\hat {\alpha}}_{\theta}({\bf 
R})$ represent the dynamics of the model in the folium ${\cal F}$. 
Furthermore, defining the space translational automorphisms ${\hat 
{\sigma}}(x)$ of ${\cal A}_{\cal F}$  by the formula
$${\hat {\sigma}}(x){\pi}(A)={\pi}\bigr({\sigma}(x)A\bigr) \ {\forall} \ 
A{\in}{\cal A}, \ x{\in}X, \eqno(3.9)$$ 
it follows from Eqs. (3.5), (3.8) and (3.9)  that the space and time 
automorphisms, ${\hat {\sigma}}(x)$ and ${\hat {\alpha}}_{\theta}(t)$, 
intercommute.
\vskip 0.5cm 
\centerline {\bf 4. The Quantum Thermodynamic Picture}
\vskip 0.3cm 
We base the quantum thermodynamic picture on 
\vskip 0.2cm\noindent
(a) a set of intercommuting, possibly unbounded,  intensive observables 
${\hat q}=({\hat q}_{0},\ {\hat q}_{1}, \ .,{\hat q}_{n})$, given by the 
global space average of ${\hat Q}({\Lambda})$ and designed to be the 
quantum counterpart of the classical variables $q$ of Section 2; and 
\vskip 0.2cm\noindent
(b) a global entropy density function, $s$, on the range, ${\cal Q}$, of  
expectation values of ${\hat q}$ in the spatially ergodic states: we term 
${\cal Q}$ the {\it thermodynamic state space}. 
\vskip 0.2cm
Specifically ${\hat q}$ is the observable at infinity [18] whose action on 
the spatially ergodic ststes given by the following formula.   
$${\hat q}({\rho})={\lim}_{{\Lambda}^{\uparrow}X}
{\vert}{\Lambda}{\vert}^{-1}{\rho}
\bigl({\hat Q}({\Lambda})\bigr){\equiv}
{\lim}_{{\Lambda}^{\uparrow}X}
{\vert}{\Lambda}{\vert}^{-1}Tr\bigl({\rho}_{\Lambda}
{\hat Q}({\Lambda})\bigr) \ {\forall} \ {\rho} \ {\in} \
{\tilde {\cal E}}_{X},\eqno(4.1)$$
${\tilde {\cal E}}_{X} $ being its domain of definition as the subset of 
${\cal E}_{X}$ on which the r.h.s. is well defined. In particular, the  
spatial ergodicity of ${\tilde {\cal E}}_{X}$ ensures that ${\hat q}$ is 
sharply defined there, i.e. that its dispersion vanishes.
\vskip 0.2cm
The construction of the entropy  function, $s$, proceeds as follows. We 
take the entropy, $S({\rho}_{\Lambda})$, of the restriction, 
${\rho}_{\Lambda}$, of the state ${\rho}$ to the region ${\Lambda} \ 
({\in}L)$,  to be given by Von Neumann\rq s formula i.e.
$$S({\rho}_{\Lambda})=- Tr_{{\cal H}({\Lambda})}
\bigl({\rho}_{\Lambda}{\rm log}({\rho}_{\Lambda})\bigr).
\eqno(4.2)$$
in units where Boltzmann\rq s constant is unity. The entropy density 
functional ${\hat s}$ of the translationally invariant states given by the 
following formula [19], which stems from the strong subadditivity 
property of $S_{\Lambda}$, established by Lieb and Ruskai [20].
$${\hat s}({\rho})={\rm lim}_{{\Lambda}{\uparrow}X}
{S({\rho}_{\Lambda})\over {\vert}{\Lambda}{\vert}} \ 
{\forall} \ {\rho} {\in}{\cal S}_{X}.\eqno(4.3)$$
Key properties of ${\hat s}$ are that it is affine and upper semi-
continuous [1]. The entropy function, $s$, on the thermodynamic space, 
${\cal Q}$ is defined by the formula 
$$s(q)={\sup}_{{\rho}{\in}{\tilde {\cal E}}_{X}}{\lbrace}
{\hat s}({\rho}){\vert}{\hat q}({\rho})=q{\rbrace}.\eqno(4.4)$$
\vskip 0.2cm
Thus the macroscopic picture provided by the quantum model is given by 
the global observable ${\hat q}$, the thermodynamic state space ${\cal 
Q}$ and the entropy density $s$ on ${\cal Q}$. Moreover, it follows 
from Eq. (4.4) and the affine property of ${\hat s}$ that the function $s$ 
is concave. Hence the tangents to the graph of $s$ at the point $q$ lie 
above that graph. Thus the set, $T_{s}(q)$, of these tangents comprises 
the control variables ${\theta}$ that satisfy the condition 
$$s(q^{\prime})-s(q){\leq}{\theta}.(q^{\prime}-q) \ {\forall} \ 
{\theta}{\in}T_{s}(q), \ q^{\prime}{\in}{\cal Q},$$
which signifies that 
$$s(q)-{\theta}.q{\geq}s(q^{\prime})-{\theta}.q^{\prime} \  {\forall} \  
q^{\prime}{\in}{\cal Q}.\eqno(4.5)$$
Consequently 
$$s(q)-{\theta}.q{\geq}{\sup}_{q^{\prime}{\in}{\cal Q}}
\bigl(s(q^{\prime})-{\theta}.q^{\prime}\bigr):={\phi}({\theta}) \ 
{\forall} \ {\theta}{\in}T_{s}(q),\ q{\in}{\cal Q}.$$
Hence
$$s(q)-{\theta}.q={\phi}({\theta}) \ {\forall} \ {\theta} 
{\in}T_{s}(q),\eqno(4.6)$$
where
$${\phi}({\theta})={\rm sup}_{q^{\prime}{\in}{\cal Q}}
\bigl(s(q^{\prime})-{\theta}.q^{\prime}\bigr).\eqno(4.7)$$
Eq. (4.6)  is just the condition for {\it global thermodynamic stability} 
(GTS), discussed in [8]. Further ${\phi}$, as defined by Eq. (4.7), is the 
reduced pressure of the quantum model [21] and is the quantum 
counterpart of the function denoted  by the same symbol for the 
phenomenological one in Section 2. Moreover, by Eq. (4.6), it is convex.
We take the control space, ${\Theta}$, to be the subset of 
${\bf R}^{n+1}$ on which ${\phi}$ is finite. As we shall infer from Eq. 
(4.15), it follows from  Eqs. (4.1)-(4.4) that this is just  
${\lbrace}{\theta}{\in}{\bf R}^{n+1}{\vert}
{\rm lim}_{{\Lambda}{\uparrow}X}{\vert}{\Lambda}{\vert}^{-1}
{\rm Tr}\bigl[{\rm exp}
\bigl(-{\theta}.{\hat Q}({\Lambda})\bigr)\bigr]<{\infty}{\rbrace}$.
\vskip 0.3cm
{\bf  The Equilibrium Condition.} We assume that the equilibrium 
condition for the model is that the entropy ${\hat s}$ is maximised, 
subject to the constraint that fixes the expectation value of ${\hat q}$. 
Moreover, as ${\hat s}$ is affine, any maximum over the states ${\cal 
S}_{X}$ must be achieved over the extremals ${\cal E}_{X}$.  Hence 
we have the following definition of thermodynamic completeness.
\vskip 0.3cm 
{\bf Definition 4.1.} {\it We term the macroscopic observables 
$Q$ thermodynamically complete if, for each $q$ in ${\cal Q}$, there is 
precisely one spatially ergodic state, ${\rho}_{q}$, that satisfies the 
equilibrium condition of maximal entropy density subject to the 
constraint that fixes the expectation value of ${\hat q}$ to be $q$. Thus, 
this constraint fixes not only the expectation value of the macroscopic 
observables ${\hat q}$ but also the full microstate of the model at the 
value ${\rho}_{q}$. Moreover, it follows from this definition that
$${\hat s}({\rho}_{q})=s(q)\eqno(4.8)$$
and
$${\hat q}({\rho}_{q})=q.\eqno(4.9)$$} 
\vskip 0.3cm
{\bf Comment.} Evidently thermodynamic completeness is a condition 
on the chosen macroscopic observables. A very simple example, which 
exhibits the significance of this condition, is provided by a ferromagnetic 
system, such as the much studied two-dimensional Ising 
model\footnote*{The argument presented here can be recast in terms of 
the more usual canonical one of Eq. (4.10) (cf. [22])}. Here the energy 
density observable ${\hat e}$ of that model is manifestly incomplete, 
since there are two spatially ergodic states with equal and opposite 
polarisation that maximise its entropy, subject to the condition that the 
expectation value $e$, of ${\hat e}$ is less than a certain critical value 
$e_{0}$. However, on supplementing the energy density ${\hat e}$ with 
a polarisation observable ${\hat m}$, one obtains a thermodynamically 
complete pair $({\hat e}, \ {\hat m})$,  for which the equilibrium state 
${\rho}_{e,m}$ is spatially ergodic.
\vskip 0.3cm
{\bf The Generalised Grand Canonical Property.} In view of Def. (4.1), 
we may re-express the formula (4.6) as the following global 
thermodynamic stability (GTS) condition [8] for the equilibrium state 
${\rho}$.
$${\hat s}({\rho})-{\theta}.{\hat q}({\rho})={\phi}({\theta}) \ {\forall} \ 
{\theta}{\in}T_{s}(q).\eqno(4.10)$$
It follows from Eqs. (4.1) and (4.2) that this condition may be expressed 
in terms of the observables $Q$ by the formula
$${\rm lim}_{{\Lambda}{\uparrow}X}{\vert}{\Lambda}{\vert}^{-1}
{\rm Tr}\bigl[{\rho}_{\Lambda}
{\rm log}{\rho}_{\Lambda}+{\rho}_{\Lambda}{\theta}.
{\hat Q}({\Lambda})\bigr]+{\phi}({\theta})=0\eqno(4.11).$$
Hence, defining the generalised canonical state ${\psi}_{\theta}$ by the 
formula
$${\psi}_{{\theta}{\Lambda}}:=
{\rm exp}\bigl(-{\theta}.{\hat Q}({\Lambda})\bigr)/{\rm Tr}
({\rm idem}) \ {\forall} \ {\Lambda}{\in}L, \eqno(4.12)$$
and introducing the relative entropy [23]
$$S({\rho}_{\Lambda}{\vert}{\psi}_{{\theta}{\Lambda}})=
{\rm Tr}\bigl({\rho}_{\Lambda}{\rm log}{\rho}_{\Lambda}-
{\rho}_{\Lambda}{\rm log}{\psi}_{\Lambda}\bigr),\eqno(4.13)$$
we infer from Eqs. (4.7) and (4.11)-(4.13) that
$${\phi}({\theta})={\rm sup}_{{\rho}{\in}{\cal E}_{X}}
{\rm lim}_{{\Lambda}{\uparrow}X}{\vert}{\Lambda}{\vert}^{-1}
\bigl[-S({\rho}_{\Lambda}{\vert}{\psi}_{{\theta}{\Lambda}})+
{\rm log}{\rm Tr}{\rm exp}(-{\theta}.{\hat Q}({\Lambda}))\bigr].
\eqno(4.14)$$ 
Consequently since $S({\xi}{\vert}{\eta})$, the entropy of a state 
${\phi}$ relative to a fixed state ${\psi}$, is minimised at the value zero 
when ${\xi}={\eta}$ [We], it follows from Eq. (4.14) thet the supremum 
on the r.h.s. of that formula is attained when 
${\rho}={\psi}_{\theta}$  and that  
$${\phi}({\theta})={\rm lim}_{{\Lambda}{\uparrow}X}
{\vert}{\Lambda}{\vert}^{-1}
{\rm log}{\rm Tr}\bigl({\rm exp}(-{\theta}.{\hat Q}({\Lambda})\bigr).
\eqno(4.15)$$
Thus, the solution of the GTS condition (4.10) is that ${\rho}$ is the 
generalised canonical state ${\psi}_{\theta}$. 
\vskip 0.3cm
{\bf Proposition 4.1.} {\it Under the above definitions and assumptions, 
the state ${\hat {\rho}}$ satisfies the KMS condtion, which may be 
expressed in the following form (cf. [HHW]).
$$\int_{\bf R}dtf(t){\langle}{\hat {\rho}};[{\hat {\alpha}}(t){\pi}_{\rho}
(A)]{\pi}_{\rho}(B){\rangle}=\int_{\bf R}dtf(t-i){\langle}
{\hat {\rho}};{\pi}_{\rho}(B)[{\hat {\alpha}}_{\theta}(t){\pi}_{\rho}(A)
{\rangle}, \ {\forall} \ A,B{\in}{\cal A}, \ f{\in}{\hat {\cal D}}({\bf R}),
\eqno(4.16)$$
where ${\hat {\cal D}}({\bf R})$ is the Fourier transform of the L. 
Schwartz space of infinitely differentiable functions on ${\bf R}$ with 
compact support.} 
\vskip 0.3cm
{\bf Proof.} By Eqs. (3.7) and (3.8) and the idntification of ${\rho}$ with 
the generalised canonical state ${\psi}_{\theta}$, the left and right sides 
of Eq. (4.16) are equal to 
$$\int_{\bf R}dtf(t){\langle}{\hat {\psi}}_{\theta};
[{\pi}_{\rho}({\alpha}_{\theta}(t)A)]{\pi}_{\rho}(B){\rangle}$$
and  
$$\int_{\bf R}dtf(t-i){\langle}{\hat {\psi}}_{\theta};
{\pi}_{\rho}(B)[{\pi}_{\rho}({\alpha}_{\theta}(t)A)]{\rangle},$$ 
respectively. Moreover, for sufficiently large ${\Lambda}, 
{\hat {\psi}}_{\theta}$ reduces to ${\hat {\psi}}_{{\theta}{\Lambda}}$ 
in these formulae. The left and right hand sides of Eq. (4.16) may 
therefore be expressed as 
$$\int_{\bf R}dtf(t){\langle}{\psi}_{{\theta}{\Lambda}};
[{\alpha}_{{\theta}{\Lambda}}(t)A)](B){\rangle}$$
and  
$$\int_{\bf R}dtf(t-i){\langle}{\psi}_{{\theta}{\Lambda}};
B[{\alpha}_{{\theta}{\Lambda}}(t)A)]{\rangle},$$ respectively. Hence,
in order to establish the formula (4.16), it suffices to show that
$$\int_{\bf R}f(t){\langle}{\psi}_{{\theta}{\Lambda}};
[{\alpha}_{{\theta}{\Lambda}}(t)A)](B){\rangle}=
\int_{\bf R}f(t-i){\langle}{\psi}_{{\theta}{\Lambda}};
B[{\alpha}_{{\theta}{\Lambda}}(t)A)]{\rangle},$$
i.e., by Eq. (4.12), that
$$\int_{\bf R}f(t){\rm Tr}
\bigl[{\rm exp}\bigl((it-1){\theta}.{\hat Q}({\Lambda})\bigr)A
{\rm exp}\bigl(-i{\theta}.{\hat Q}({\Lambda})\bigr)B\bigr]=$$
$$\int_{\bf R}f(t-i){\rm Tr}
\bigl[{\rm exp}\bigl(-{\theta}.{\hat Q}({\Lambda})\bigr)
B{\rm exp}\bigl(i{\theta}.{\hat Q}({\Lambda})t\bigr)
A{\rm exp}\bigl(-i{\theta}.{\hat Q}({\Lambda})t\bigr)\bigr].
\eqno(4.17)$$
By the change of variable from $t$ to $(t-i)$, the l.h.s. takes the form
$$\int_{\bf R}f(t-i){\rm Tr}\bigl[{\rm exp}
\bigl(i{\theta}.{\hat Q}({\Lambda})t\bigr)A{\rm exp}\bigl(-i{\theta}.
{\hat Q}({\Lambda})(t-i)\bigr)B\bigr];$$
and, in view of the cyclicity of ${\rm Trace}$, this is equal to
$$\int_{\bf R}f(t-i){\rm Tr}\bigl[B{\rm exp}
\bigl(i{\theta}.{\hat Q}({\Lambda})t\bigr)A{\rm exp}\bigl(-i{\theta}.
{\hat Q}({\Lambda})t\bigr)
{\exp}\bigl(-{\theta}.{\hat Q}({\Lambda})\bigr)\bigr],$$
which, again by the cyclicity of ${\rm Trace}$, is equal to the r.h.s. of 
Eq. (4.17), as required.
\vskip 0.5cm 
\centerline {\bf 5. Differentiability of Entropy}
\vskip 0.3cm
The condition for differentiability of the entropy function $s$ is simply 
that the tangent space, $T_{s}(q)$ at each point $q$ of ${\cal Q}$ is one-
dimensional.  We shall now  relate this condition to the dynamics of the 
model, subject to the following assumption, that appears to be natural in 
view of the definition (3.8) of the dynamical automorphisms 
${\hat {\alpha}}_{\theta}({\bf R})$.
\vskip 0.3cm
{\bf Assumption 5.1.} {\it ${\hat {\alpha}}_{{\theta}_{1}}(t)=
{\hat {\alpha}}_{{\theta}_{2}}(t) \ {\forall} \ t{\in}{\bf R}$ if and only 
if ${\theta}_{1}={\theta}_{2}$. }
\vskip 0.3cm
{\bf Proposition 5.1} {\it Under the Assumptions 5.1, the entropy 
function $s$ is differentiable.} 
\vskip 0.3cm
{\bf Proof.} Assume that $s$ is not differentiable at the point $q$. Then 
the tangent space $T_{s}(q)$ contains at least two different elements, 
${\theta}_{1}$ and ${\theta}_{2}$, of ${\Theta}$. Hence, by Prop. 4.1, 
the state ${\rho}$ satisfies the KMS condition with respect to the 
automorphism groups  ${\hat {\alpha}}_{{\theta}_{1}}$ and 
${\hat {\alpha}}_{{\theta}_{2}}$. On the other hand, the thermodynamic 
completeness of the observables $Q$ ensures the uniqueness of the 
spatially ergodic equilibrium state ${\hat {\rho}}_{q}$ under the 
prevailing conditions. Therefore this state must satisfy the KMS 
conditions with respect to the automorphisms ${\hat 
{\alpha}}_{{\theta}_{1}}$ and 
${\hat {\alpha}}_{{\theta}_{2}}$, and consequently, by the uniquness of 
the modular automorphisms [2], the groups 
${\hat {\alpha}}_{{\theta}_{1}}({\bf R})$ and 
${\hat {\alpha}}_{{\theta}_{2}}({\bf R})$ must coincide. Hence, by 
Assumption 5.1, ${\theta}_{1}$ and ${\theta}_{2}$ must be equal, 
which is contrary to hypothesis. We conclude that $T_{s}(q)$ must be 
one-dimensional and consequently, as $q$ is an arbirary point of 
${\cal Q}$, that $s$ is differentiable. 
\vskip 0.5cm
\centerline {\bf  6. Concluding Remarks}
\vskip 0.3cm
An aim of this article has been to present a model independent approach 
to quantum thermodynamics, based on the assumptions of 
thermodynamic completeness, defined in Def. 4.1, and the condition 
(3.6), whereby the folium ${\cal F}$ supports a dynamics  given by the 
infinite volume limit of that of a corresponding finite system. These 
assumptions serve to extend the quantum thermodynamic picture from 
the lattice systems, on which much of the constructive work on the 
subject is based [6, 15-18], to continuous ones. Further, the resultant 
dynamics is given by the automorphisms  
${\hat {\alpha}}_{\theta}({\bf R})$, dual to ${\cal F}$, defined by Eq. 
(3.8).
\vskip 0.2cm 
We note here that, apart from its intrinsic significance, the completeness 
condition plays a key role in the derivation of the differentiability of 
entropy, which, in turn, represents consistency between the quantum and 
phenomenological pictures. This consistency appears to be nontrivial, 
since, in the phenomenological picture, this differentiability is just 
Caratheodory\rq s theorem, whereas in the quantum picture, it arises from 
Von Neumann\rq s formula (4.1) for the entropy, which appears to be  
quite different.
\vskip 0.2cm
Finally, we remark that the theory presented here is centred on the 
entropy function, $s$, whereas the previous works on quantum 
thermodynamics are based largely on its Legendre transform ${\phi}$, 
which is just the reduced pressure function. A radical difference between 
the properties of $s$ and ${\phi}$ is that, while the former is 
differentiable, the latter is generally not so, as phase transitions occur at 
just those values of the control variables where the  reduced pressure is 
not differentiable [1].
\vskip 0.5cm
\centerline {\bf References}
\vskip 0.3cm\noindent
[1] D. Ruelle: {\it Statistical Mechanics}, W. A. Benjamin, New York, 
1969
\vskip 0.2cm\noindent
[2] M. Takesaki:  {\it Tomita\rq s Theory of Modular Hilbert Algebras 
and its Applcations. Lecture Notes in Mathematics Vol. 128. Springer, 
Berlin 1970}
\vskip 0.2cm\noindent
[3] H. Callen: {\it Thermodynamics and an Introduction to 
Thermostatistics}, Wiley, New York, 1985
\vskip 0.2cm\noindent
[4] R. Haag, N. M. Hugenholtz and M. Winnink: Commun. Math. Phys. 
{\bf  5}, 215-236, 1967
\vskip 0.2cm\noindent
[5] G. G. Emch: {\it Algebraic Methods in Statistical Mechanics and 
Quantum Field Theory}, Wiley, New York, 1972
\vskip 0.2cm\noindent
[6] O. Bratteli and D. W. Robinson: {\it Operator Algebras and Statistical 
Mechanics II}, Springer, Heidelberg, 1981
\vskip 0.2cm\noindent
[7] W. Thirring: {\it Quantum Mechanics of Large Systems},  Springer, 
New York, 1983
\vskip 0.2cm\noindent
[8] G. L. Sewell: {\it Quantum Mechanics and its Emergent 
Macrophysics},  Princeton University Press, Princeton, 1982
\vskip 0.2cm\noindent
[9] G. L. Sewell: J. Math. Phys. {\bf 11}, 1868-1884, 1970
\vskip 0.2cm\noindent
[10]  G.-F. Dell\rq Antonio, S. Doplicher and D. Ruelle: Commun. Math. 
Phys. {\bf 2}, 223-230, 1966
\vskip 0.2cm\noindent
[11] R. Haag, R. V. Kadison and D. Kastler: Commun. Math. Phys. 
{\bf  16}, 81-104, 1970
\vskip 0.2cm\noindent
[12] ]D. A. Dubin and G. L. Sewell: J. Math. Phys. {\bf 11}, 2990-2998, 
1970
\vskip 0.2cm\noindent
[13] G. L. Sewell: Lett. Math. Phys. {\bf 6}, 209-213, 1982
\vskip 0.2cm\noindent
[14] R. F. Streater: Commun. Math. Phys. {\bf  6}, 233-247, 1967
\vskip 0.2cm\noindent
[15] D. W. Robinson: Commun. Math. Phys. {\bf 6}, 151-160, 1967
\vskip 0.2cm\noindent
[16] D. W. Robinson: Commun. Math. Phys. {\bf 7}, 337-348, 1968
\vskip 0.2cm\noindent
[17] O. E. Lanford and D. W. Robinson: Commun. Math. Phys. {\bf  9}, 
327-338, 1968
\vskip 0.2cm\noindent
[18] O. E. Lanford and D. Ruelle: Commun. Math. Phys. {\bf 13}, 194-
215, 1969
\vskip 0.2cm\noindent
[19] H. Araki and E. H. Lieb: Commun. Math. Phys. {\bf 18}, 160-170, 
1970
\vskip 0.2cm\noindent
[20] E. H. Lieb and M. B. Ruskai: J. Math. Phys. {\bf 14}, 1938-1941, 
1973
\vskip 0.2cm\noindent
[21] D. W. Robinson: {\it The Thermodynamic Pressure in Quantum 
Statistical Mechanics}, Lec. Notes in Physics, Vol. 9,, Springer, Berlin 
1971
\vskip 0.2cm\noindent
[22] A. Messager and S. Miracle-Sole: Commun. Math. Phys. {\bf 40}, 
187-198, 1975
\vskip 0.2cm\noindent
[23] A. Wehrl: Rev. Mod. Phys.{\bf 50}, 221-260, 1978

\end